\documentclass[cornellheadings,smallerheadings]{interdepnetworks}
\usepackage{amsmath} 
\usepackage{amssymb} 
\usepackage{graphicx} 
\usepackage{amsthm}
\usepackage{color}
\usepackage{makeidx}
\usepackage{mathtools}
\usepackage{enumerate}
\usepackage[english]{babel}
\usepackage{tikz}
\usepackage{lettrine}
\usepackage{pgfplots}
\usepackage{scrextend}
\usepackage{caption}
\usepackage{subcaption}
\usepackage{algorithm}
\usepackage{algorithmic}
\usepackage{multirow}
\usepackage{hhline}
\usepackage{colortbl}
\usepackage{rotating}
\usepackage{comment}

\usepackage[withpage]{acronym}

\usetikzlibrary{arrows,shapes,snakes,calc,fit,intersections,patterns}

\title{Control of Cooperative Unmanned Aerial Vehicles: Review of Applications, Challenges, and Algorithms}

\chapterid{TBD}
\author{Arman Sargolzaei $^{1,*}$, Alireza Abbaspour  $^{2}$, Carl D. Crane$^{1}$}

\affiliationone{(1) Department of Mechanical and Aerospace Engineering, University of Florida, Gainesville, FL, USA.\\}
\affiliationtwo{(2) Department of Advanced Engineering, Hyundai Mobis, Plymouth, Michigan, USA.\\}
\emailaddresses{a.sargolzaei@gmail.com, aabbaspour@mobis-usa.com, ccrane@ufl.edu}

\chapterabstract{A system of cooperative unmanned aerial vehicles (UAVs) is a group of agents interacting with each other and the surrounding environment to achieve a specific task. In contrast with a single UAV, UAV swarms are expected to benefit efficiency, flexibility, accuracy, robustness, and reliability. However, the provision of external communications potentially exposes them to an additional layer of faults, failures, uncertainties, and cyber-attacks and can contribute to the propagation of error from one component to other components in a network. Also, other challenges such as complex nonlinear dynamic of UAVs, collision avoidance, velocity matching, and cohesion should be addressed adequately. Main applications of cooperative UAVs are border patrol; search and rescue; surveillance; mapping; military. Challenges to be addressed in decision and control in cooperative systems may include the complex nonlinear dynamic of UAVs, collision avoidance, velocity matching, and cohesion. In this paper, emerging topics in the field of cooperative UAVs control and their associated practical approaches are reviewed.}


\begin{document} 
\setcounter{page}{1}
\pagenumbering{roman}
\pagestyle{plain}

\contentspage
\normalspacing
\setcounter{page}{1}
\pagenumbering{arabic}
\pagestyle{cornell}
\makechaptertitle

\section*{Abbreviations}
\begin{acronym}[JSONP]\itemsep-8pt
\acro{UAV}{Unmanned Aerial Vehicles}
\acro{ROS}{Robot Operating System}
\acro{CCUAVs}{Cooperative Control Unmanned Aerial Vehicles}
\acro{POMDP}{Observable Markov Decision Process}
\acro{LIDAR}{light detection and ranging}
\acro{CML}{concurrent mapping and localization}
\acro{DDF}{Decentralized Data Fusion}
\acro{SLAM}{Simultaneous Localization and Mapping}
\acro{CLSF}{Constrained Local Submap Filter}
\acro{PRS}{Personal Remote Sensing}
\acro{SWEEP}{Swarm Experimentation and Evaluation Platform}
\acro{ECM}{Electronic Counter-Measure}
\acro{UCAVs}{Unmanned Combat Air Vehicles}
\acro{EJ}{Escort Jamming}
\acro{SAM}{Surface to Air Missile}
\acro{FTC}{Fault Tolerant Controllers}
\acro{FDI}{Fault Detection and Identification}
\acro{LQR}{Linear Quadratic Regulator}
\acro{DDDAS}{Dynamic Data-Driven Application System}
\acro{PDF}{Probability Density Function}
\acro{PP}{Pure pursuit}
\acro{LOS}{Line of Sight}
\acro{PN}{Proportional Navigation}
\acro{DI}{Dynamic Inversion}
\acro{GNN}{Grossberg Neural Network}
\acro{WSN}{Wireless Sensor Network}
\acro{PC}{probability collective}
\acro{TDS}{Time delay switch}
\acro{DoS}{Denial of Service}
\end{acronym}
\section{Introduction}
Swarm intelligence deals with physical and artificial systems formed of entities that have internal and external interactions coordinating by incentive or a predefined control algorithm. Flocking of birds, swarming of insects, shoaling of fishes, and herding of quadrupeds,  were a motive for the cooperated control of UAVs. A group of UAVs can be modeled similar to natural animal cooperation where bodies operate as a system toward reaching mutual benefits. Animals can benefit from swarm performance in defending against predators, food seeking, navigation, and energy saving. Cooperative multi-robots complete a task in a shorter time \cite{lin2010supporting}, have synergy \cite{riehl2011cooperative,how2004flight}, and cover a larger area. They are also more cost-effective using smaller, simpler, and more durable robots \cite{hung2017q}. Furthermore, they can complete a task more accurately and robustly \cite{mccune2014control}. 

Cooperative control is one of the most attractive topics in the field of control systems which has received the attention of many researchers. Cooperative algorithms and utilization are mainly discussed in the recent decade. Many useful surveys have been done to review the recent contributions in this field \cite{ren2007information,anderson2008rigid,wang2014overview,oh2015survey,zhu2017survey,senanayake2016search}. However, most of them were just focused on the algorithms and on the consensus control theory. A valuable review of the consensus control problem was done by Ren et al. \cite{ren2007information}; however, significant contributions have been done thereafter. Anderson et al.  \cite{anderson2008rigid} also focused on consensus control of the multi-agent systems. Wang et al. \cite{wang2014overview} and Zhu et al. \cite{zhu2017survey} reviewed most of the consensus control problems, however, other cooperative techniques and the application of these algorithms were not discussed. Senanyake et al. investigated cooperative algorithms for searching and tracking applications \cite{senanayake2016search}; however, the other algorithms and applications of the cooperative system were not considered.

To enhance the current related literature mentioned above and cover most of the applications and algorithms, the recent research studies in the field of cooperative control design will be reviewed. The applications, algorithms, and challenges are considered. The applications are categorized into surveillance, search and rescue, mapping, and military applications; then, the recent developments related to each category are reviewed. Similarly, the algorithms can be categorized into three main classes: consensus control, flocking control, and guidance based cooperative control. The challenges related to the cooperative control and applications of cooperative algorithms are investigated in a separate section. Moreover, the related mathematics of cooperative control algorithms are simplified to make it easier for readers to understand the concepts.\\
This paper is organized as follows: Section 2 provides potential applications of cooperative control, and Section 3 highlights possible challenges when applying cooperative control. Section 4 reviews  algorithms used in cooperative control design. Finally, Section 5 provides the summary and conclusion of this work.
\section{Applications and literature review}
Cooperative control of multiple unmanned vehicles is one of the topics in control areas that have received increasing interest in the past several years. Single UAVs have been applied for various applications, and recently, investigators have attempted to expand and improve their applications by using a combination of multiple agents. The multiple agents concept has been used for search and rescue \cite{goodrich2007using,riehl2011cooperative,cui2015drones,nigam2008control,nigam2012control,ahmadzadeh2006multi}, geographic mapping \cite{fenwick2002cooperative,remondino2011uav,lin2011mini,zongjian2008uav}, military applications \cite{mears2005cooperative,junwei2014target,paley2008cooperative}, etc. In this section, the current and potential applications of the cooperative control of UAVs are surveyed.

\subsection{Search and Rescue}
\begin{figure} [!h]
\begin{center}
\includegraphics[scale=0.5]{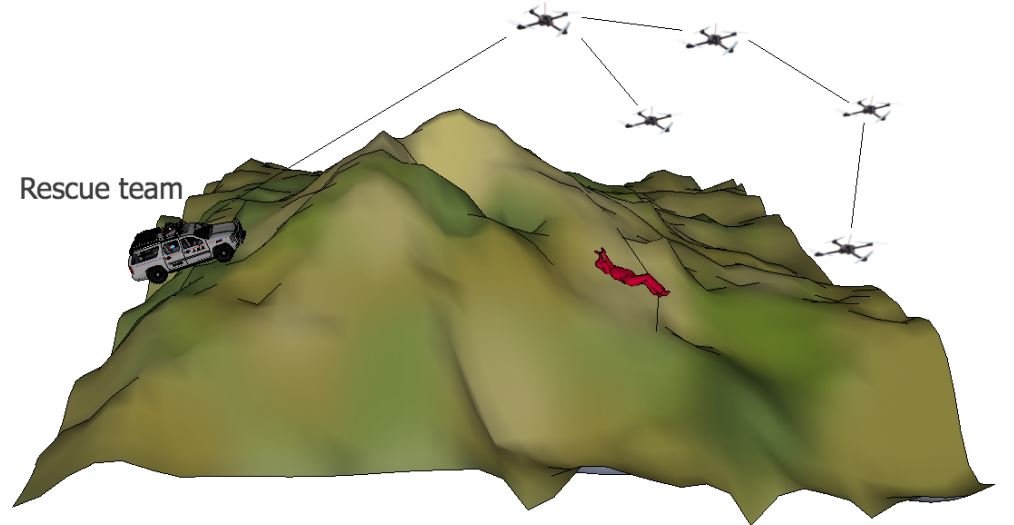} 
\caption {A cooperative approach to search for a victim in a hard to access area \cite{waharte2010supporting}  }
\label{F:Search_Rescue}
\end{center}
\end{figure}

UAVs have been used for several years for search and rescue purposes since they are more compact and cost-effective and require less amount of time to deploy than a plane or helicopter, particularly when multiple numbers of UAVs are required to accomplish the task. Figure \ref{F:Search_Rescue} displays a scenario for cooperative control of quadrotors to search for and rescue a patient or missing person in a hard to access environment. 
In this kind of operation, search time is the most critical factor. To satisfy the time constraint, Scherer et al. implemented a distributed control system in the Robot Operating System (ROS) of the multiple multi-copters, to capture the situations and display them as video streams in real-time at base stations  \cite{scherer2015autonomous}. Since UAVs have their advantages such as agility, swiftness, remote-controlling, bird’s eye-vision, and other integrities, they can creditably perform practical work promptly. However, when those advantageous of UAVs are operated by a cooperative control algorithm to complete a mission, the requirement of minimum time-delay and other critical constraints can be achieved in searching and rescuing casualties or victims. 

Many types of research and experiments are performed in search and rescue requiring Cooperative Control Unmanned Aerial Vehicles (CCUAVs). For example, Waharte et al. showed that employing multiple autonomous UAVs has excellent benefits in the search and rescue operations for the corollary of Hurricane Katrina in September 2006. The notable sophistication of their work was that they divided the real-time approaches into three main categories which were Greedy heuristics, Potential-based heuristics, and Partially Observable Markov Decision Process (POMDP) based heuristics \cite{waharte2010supporting}. 
In a case of fire,  Maza et al. investigated a multi-UAV firefighters monitoring mission in the framework of the AWARE Project using two autonomous helicopters to monitor the firemen's performance and safety in real time from a simulated situation where firefighters are assisting injured people in front of a burning building. This work has been done based on their previous work's algorithm \cite{viguria2010distributed}, SIT algorithm, which follows a market-based approach combined with a network of ground cameras and a Wireless Sensor Network (WSN) \cite{maza2010firemen}. 
Another scenario that CCUAVs can be wholly beneficial is to search and rescue missing persons in a wilderness. It has been many centuries that travelers had been lost in wildernesses such as mountains, oceans, deserts, jungles, rain forests, or any abandoned or uncolonized areas. Some of the missing people could be found and rescued, but many of them were lost from their families forever. Goodrich et al. have shown and identified a set of operational practices for using mini Unmanned Aerial Vehicles (mUAVs) to support wilderness search and rescue (WiSAR) operations. In their work, technical operations such as sequential operations, remote-led operations, and base-led operations have been used to gather and analyze evidence or potential signs of a lost person to simulate a stochastic model of his behavior and a geographic description of a particular region. If the model is well matched to a specific victim, then the location of the missing person would be estimated according to the probability of the area where the lost person could be located \cite{goodrich2007using}. The result of their research shows that the mUAVs could address the limitations of human-crewed aircraft which also upholds the research algorithm of CCUAVs.

\subsection{Surveillance}
\begin{figure} [!h]
\begin{center}
\includegraphics[scale=0.5]{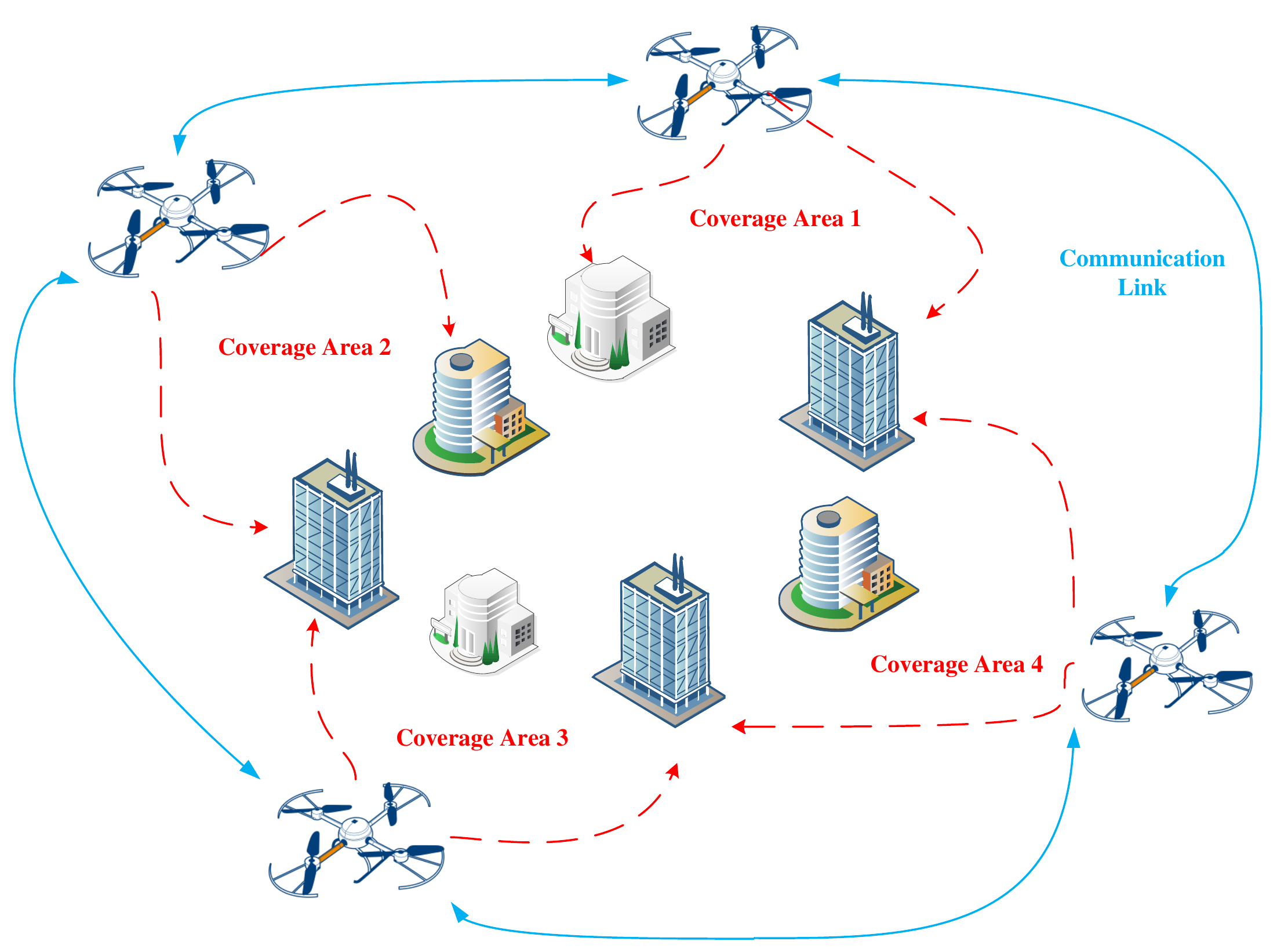} 
\caption {Cooperative surveillance concept for patrolling urban areas.}
\label{F:Survillence}
\end{center}
\end{figure}

Surveillance is one of the applications of UAVs that have been widely used. Figure \ref{F:Survillence} shows an overall scheme of the surveillance application using the cooperative quadrotors system. \\
Bread et al. studied aerial surveillance of fixed-wing multi-UAVs. Fixed-wing aircraft may have a significant advantage in speed. However, the lack of hovering ability would increase their chance of collision when they work in cooperative control mode. To mitigate and overcome this constraint, Bread et al. presented an approach which consists of four significant steps: cooperation objective and constraints, coordination variable and coordination function, centralized cooperation scheme, and consensus building \cite{beard2006decentralized}.

Ahmadzadeh et al. \cite{ahmadzadeh2006multi} have studied the cooperative motion-planning problem for a group of heterogeneous UAVs. In their work, the surveillance operations were conducted via the body-fixed cameras equipped on their fixed-wing UAV. They demonstrated multi-UAV cooperative surveillance with spatiotemporal specifications \cite{ahmadzadeh2006multi}. Besides, they used an integer programming strategy to reduce the computational effort. The main contribution of their study was to generate an appropriate trajectory associated with the complexities of coupling cameras field of view with flight paths. 
Paley et al. designed a glider with a coordinated control system for long-duration ocean sampling using real-time feedback control \cite{paley2008cooperative}. In their design, agents were modeled as Newtonian particles to steer a set of coordinated trajectories. However, this model cannot be applied for closed flocking due to the assumption that there is enough space between particles.

In the case of persistent surveillance, Nigam et al. have intensively researched on UAVs for persistent surveillance and their works have been consecutively released in the past few years. Their early efforts focused on investigating techniques for a high-level, scalable, reliable, efficient, and robust control of multiple UAVs \cite{nigam2008control} and derived an optimum policy with a single UAV \cite{nigam2008persistent}. They also suggested that modifications of the existing control policies would improve the system performance under dynamic constraints and proposed multi-agent reactive policy to integrate multiple UAVs and optimized the performance using a real-encode probability collective (PC) optimization framework. In the later works, Nigam et al. have developed algorithms to control multiple UAVs for persistent surveillance and devised a semi-heuristic approach for a surveillance task using multiple UAVs \cite{nigam2012control}. Their research considered the effect of aircraft dynamics on the performance of the designed cooperative mission and the advantages of their policy's performance was demonstrated by comparing it with other benchmark approaches such as the potential field-like approach, the planning-based approach, and the optimum approach.
Paley and Peterson developed their previous research for ocean sampling \cite{paley2008cooperative}, for environmental monitoring and surveillance \cite{paley2009stabilization}. Each UAV was considered as a Newton particle which was incorporated in a gyroscopic steering control system. This design has several drawbacks: first, obstacle avoidance in Newton particle method is not considered; second, all UAVs are moving in the same direction which is not flexible for surveillance and searching tasks; third, each UAV orbit around an inertially fixed point at constant radius which is not an energy efficient method for monitoring and surveillance.
\subsection{Localization and mapping}
\begin{figure} [!h]
\begin{center}
\includegraphics[scale=0.5]{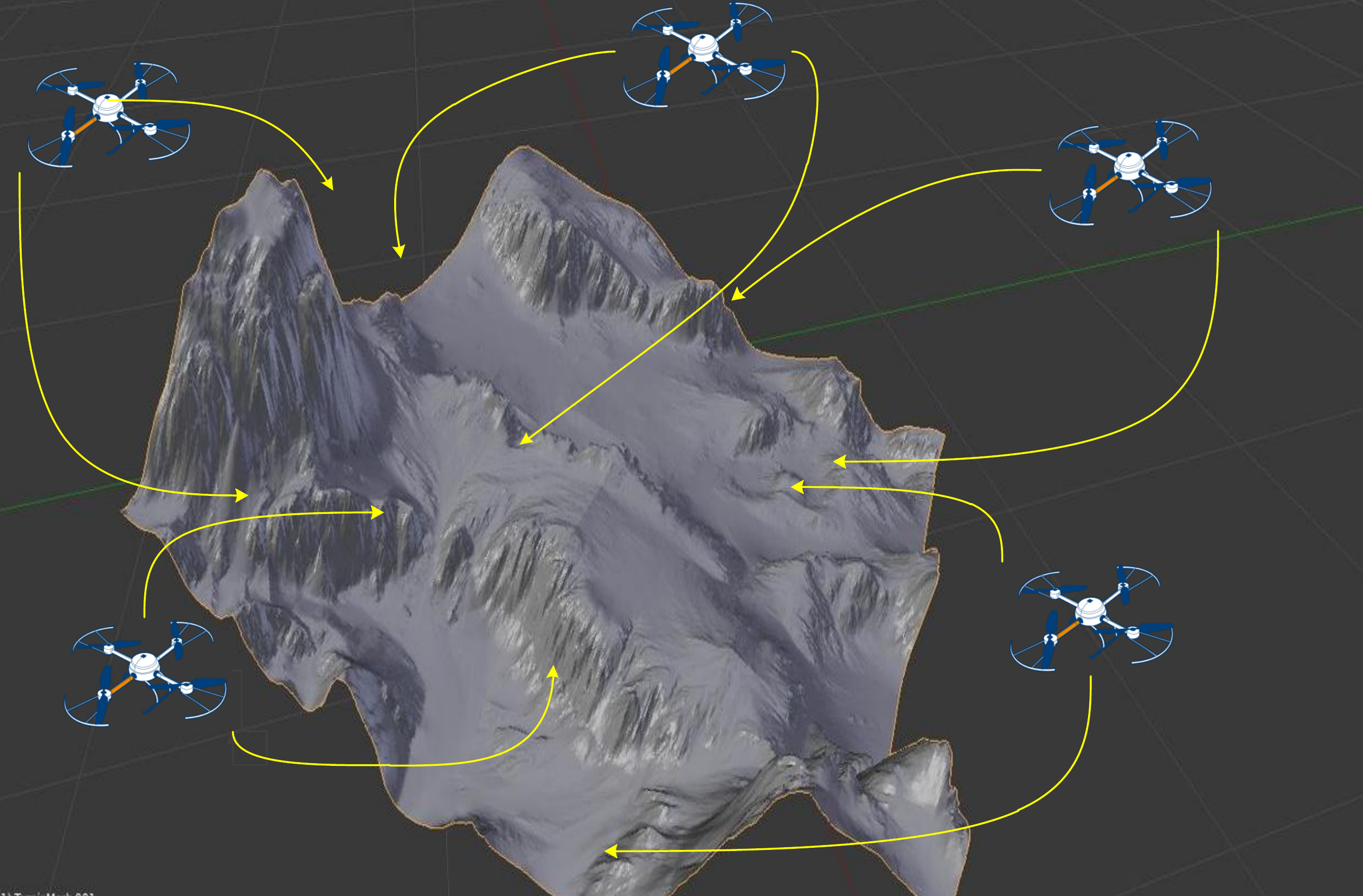} 
\caption {Cooperative three-dimensional mapping using quadrotor UAVs.}
\label{F:Mapping}
\end{center}
\end{figure}

High agility, wide vision, and accessibility are some of the significant factors that made the UAVs a popular tool to map and model lands or terrains \cite{remondino2011uav}. UAVs have been used to map in several types of research \cite{remondino2011uav,lin2011mini,zongjian2008uav}. Figure \ref{F:Mapping} shows the concept of cooperative 3D mapping by multiple quadrotors.  Remondino et al. used UAVs for space-mapping and 3D-modeling in several types of vehicles and techniques \cite{remondino2011uav}. 
One of the high systems in the mapping technology of UAVs is known as light detection and ranging (LIDAR) was employed by Lin et al. \cite{lin2011mini}. They have applied the LIDAR-based system on a mini-UAV-borne cooperating with Ibeo Lux and Sick laser scanners and an AVT Pike F-421 CCCD camera to map a local area in Vanttila, Espoo, Finland in a fine-scale. 

As the surveillance and searching algorithms, the cooperative mapping task of UAVs can help to improve the accuracy and reduce the operation time through sharing their responsibilities. Cooperative control of autonomous vehicles can be used to make a map for an unknown environment and 3D-modeling. Fenwick et al. introduced a novel algorithm for concurrent mapping and localization (CML) which combines the information of navigation and sensors of multiple unmanned vehicles \cite{fenwick2002cooperative}. This algorithm is working based on stochastic estimation and to extract landmarks from the mapping area using a feature-based approach. Goktogan et al. developed and demonstrated the multiple sensing nodes of numerous UAV platforms using Decentralized Data Fusion (DDF) algorithm to simultaneously localize and map the flight simulator in real time \cite{goktogan2003real}.

Simultaneous Localization and Mapping (SLAM) presented by Williams et al. \cite{williams2002towards} can be used to examine the prospect of the Constrained Local Submap Filter (CLSF) algorithm and applied to the multi-UAVs as SLAM algorithm. The advantage of this approach is that it allows the cross-covariance process to be scheduled at convenient intervals and aids in the data association problem.

Localization and mapping in unsafe or obscure places is another critical application of UAVs. Multi-UAV cooperative control has been used for mapping in wild or unknown areas in several types of research  \cite{bryson2007co,han2013low} such as the continuation of the SLAM algorithm and its applications presented by Bryson and Sukkarieh \cite{bryson2007co}. Han et al. have introduced Personal Remote Sensing (PRS) multi-UAVs for contour mapping in two scenarios of nuclear radiation \cite{han2013low}. Their work also focused on the costs of the multi-UAVs and the efficiency of atomic radiation detection in a necessary time which were the main advantages over a single UAV mapping. Kovacina et al. also focused on mapping a hazardous substance which was a chemical cloud. To map the chemical cloud, Kovacina et al. used Swarm Experimentation and Evaluation Platform (SWEEP) with their developed rule-based, decentralized control algorithm to simulate an air vehicle swarm searching for and mapping a chemical cloud \cite{kovacina2002multi}.

\subsection{Military Applications}
The cooperative control of UAVs has various practical and potential military application varies from reconnaissance and radar deception to surface-to-air-missile jamming. It has been demonstrated that a group of low-cost and well-organized UAVs can have better effects than a single high-cost UAV \cite{jeon2010homing}. Generally, the application of cooperative control for the unmanned system in the military can be categorized into two main categories: reconnaissance and penetrating strategies. To achieve these types of applications, UAVs may need to flying near each other with a specific structure. Formation flight control is one of the most straightforward cooperative strategies which consists of a set of aircraft flying near to each other in a defined distance \cite{sadeghi2015novel}. One of the advantages of flight formation is a significant reduction in fuel consumption through locating the follower aircraft such that the vortex of the leader aircraft reduces the induced drag of the follower aircraft \cite{lavretsky2002f}.
\subsubsection{Reconnaissance Strategy}
\begin{figure} [!h]
\begin{center}
\includegraphics[scale=0.4]{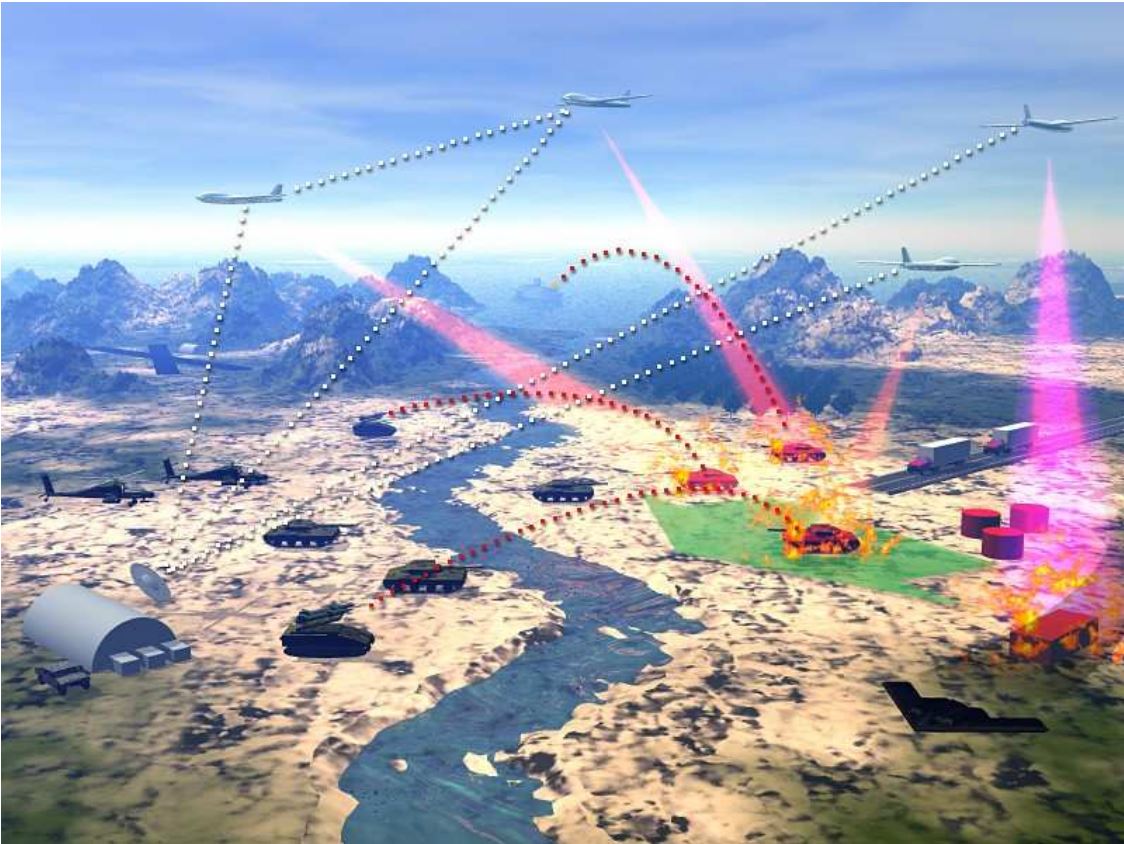} 
\caption {Target detection using multiple cooperative UAVs in a reconnaissance mission \cite{murray2007recent}.}
\label{F:4}
\end{center}
\end{figure}

A formation or cooperative design of UAVs can be used as reliable radars or reconnaissance tools to detect enemy troops and ballistic missiles \cite{liu2012ballistic, stone2000radars}. The integration of the UAVs radars will help to identify incursion objects or observe ground activities of an adversary \cite{schwartz1990radar,wang2014cooperative}. Ahmadzadeh et al. introduced a cooperative strategy to enable a heterogeneous team of UAVs to gather information for situational awareness \cite{ahmadzadeh2006cooperative}. In their work, an overall framework for reconnaissance and an algorithm for cooperative control of UAVs considering collision and obstacle avoidance were presented. Figure \ref{F:4} shows a reconnaissance mission using multiple cooperative UAVs.
\subsubsection{Penetrating Strategy}
The new and robust defense mechanism of rivals makes it difficult to penetrate to their territories. To this aim, various strategies have been designed to deceive the target radar and defense mechanism \cite{kim2004cooperative, lee2006guidance,jeon2006impact}. 

\begin{figure} [!h]
\begin{center}
\includegraphics[scale=0.4]{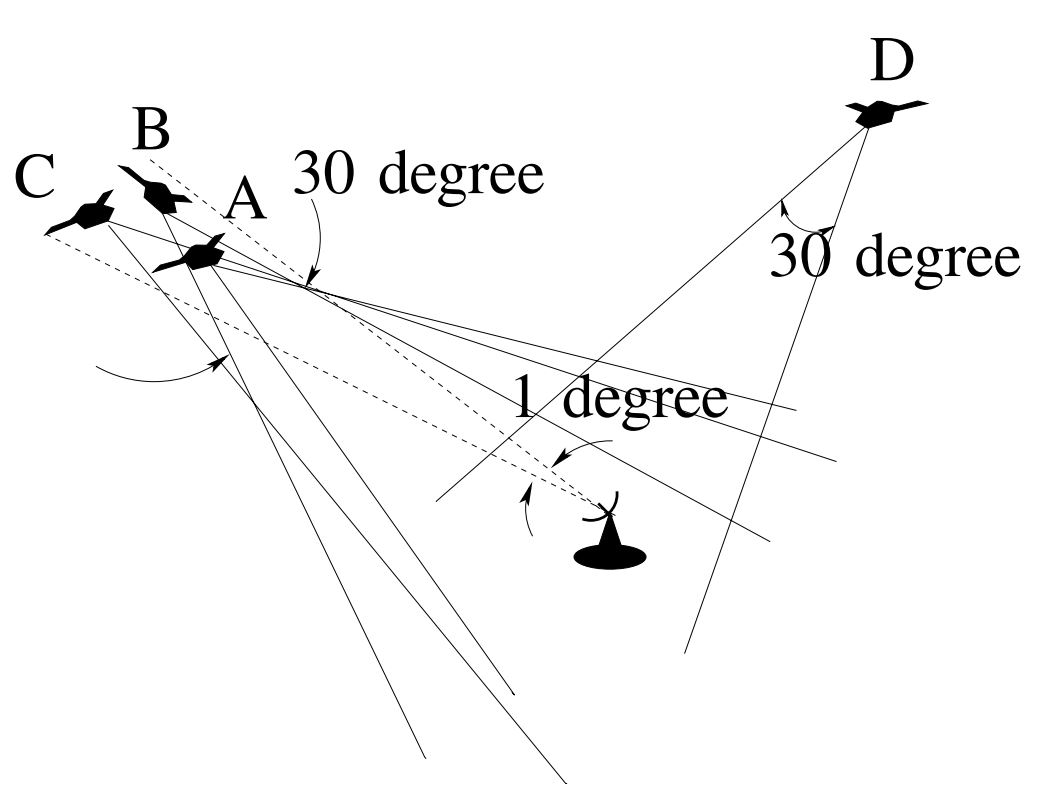} 
\caption {Cooperative radar jamming using multiple UAVs \cite{kim2004cooperative}.}
\label{F:5}
\end{center}
\end{figure}

Being hidden from the enemy radars through Electronic Counter-Measure (ECM) is called Radar jamming which is a very important action that is mostly used by Unmanned Combat Air Vehicles (UCAVs) to protect or defend themselves from surface-to-air missiles when the vehicles reconnoiter into enemy territories. The radar jamming consists of sending some noise to deceive the enemies radar signal. The radar jamming and deception can be more effective when a group of UCAVs works together. Jongrae et al. focused on the Escort Jamming (EJ) of the UAVs while a close formation and cooperative control procedure are designed to deceive the tracking radar of the Surface to Air Missile (SAM) \cite{kim2004cooperative}. Generally, jamming can be classified into two categories: self or support jamming. Figure \ref{F:5} shows the two mentioned methods of interference, where "D" shows the self-jamming and "A, B, C" UAVs show the support jamming.   

\begin{figure} [!h]
\begin{center}
\includegraphics[scale=0.4]{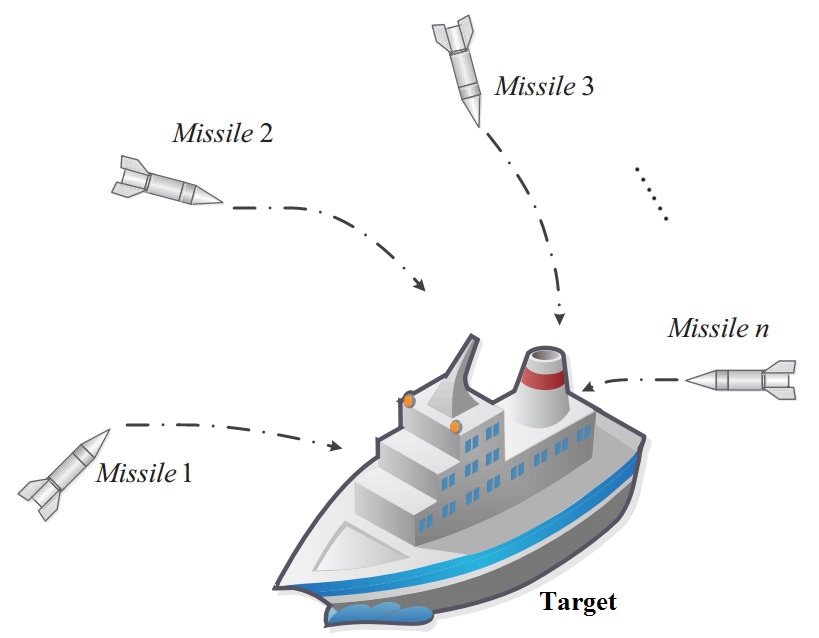} 
\caption {Cooperative missile attacking  concept to a target.}
\label{F:6}
\end{center}
\end{figure}

The missiles control system is similar to the UAV control system. However, they are not designed to come back to the station. Since penetrating to the high-tech defense mechanism of a target is very complicated, a group of cooperative missiles will have more chance to penetrate a defense mechanism in comparison with being independently operated \cite{lee2006guidance,jeon2006impact}.  Figure \ref{F:6} shows a collective missile attack to a ship target.\\

\section{Challenges}

Multi-UAV systems have advantages over single UAVs in the impact of failure, scalability, survivability, the speed of the mission, cost, required bandwidth, and range of antennas \cite{gupta2016survey}. However, these systems are complex and hard to coordinate. Gupta and Vaszkun considered three challenges in providing a stable and reliable UAV network: architectural design of networks; routing the packet from an origin to a destination and optimizing the metric; transferring from an out-of-service UAV to an active UAV, and energy conservation \cite{gupta2016survey}. 
According to a study at MIT, the main challenges associated with the development and testing of cooperative UAVs in dynamic and uncertain situations are real-time planning; designing a robust controller; and using communication networks \cite{paley2008cooperative}.
Ryan et al. address issues in cooperative UAV control which are: aerial surveillance, detection, and tracking which allows vision-based control; collision and obstacle avoidance and formation reconfiguration; high-level control needed for real-time human interfacing; and security of communication links \cite{ryan2004overview}. Oh, et al. addressed the problem of modeling the agent's interactions with each other and with the environment which is challenging to predict \cite{oh2017bio}. The most significant challenges in cooperative control of multi-agent systems can be summarized as below.\\
\textbf{1)} In cooperative control, instead of developing a control objective for a single system, it is necessary to devise control objectives for several subsystems. Moreover, the relation between the team goal and agent goal needs to be negotiated and balanced \cite{lewis2013cooperative}.\\
\textbf{2)} The communication bandwidth and quality of connection among agents in the system are limited and variable. Moreover, the security of communication links in the presence of intruders should be considered in the design \cite{martini2015distributed,khaldi2017monitoring,abbaspour2017adaptive,abbaspour2016detection}. The CUAV is vulnerable to a range of cyber attacks such as Denial of Service (DoS) and time delay switch (TDS) attacks \cite{sargolzaei2015preventing,sargolzaei2014delayed,sargolzaei2017resilient,sargolzaei2018security}. \\
\textbf{3)} The aerodynamic interference of the agents on each other should be considered in the design \cite{oh2017bio}. Close cooperative flight control or formation has also specific aerodynamic challenges which are called aerodynamic coupling. These aerodynamic interferences are caused by the vortex effect of the leading aircraft and should be modeled and quantified in the controller design to avoid their critical impact on the system stability. Otherwise, unwanted rolling or yawing moment will be generated which can destabilize the overall system \cite{giulietti2005dynamic,pereira2016tight}. However, incorporating the coupled dynamic in the formation design can help to reduce energy consumption through the mission \cite{ray2002flight,pahle2012preliminary}.\\
\textbf{4)} The controller design of CUAV should include fault tolerable algorithms through software redundancy because hardware redundancy is not an option for mini-UAVs. The fault tolerant control design for one UAV is a challenging task by itself, which has been discussed in the literature \cite{abbaspour2018neural,abbaspour2017neural}. 
\section{Algorithms}
\begin{figure}
\begin{center}
\includegraphics[scale=0.75]{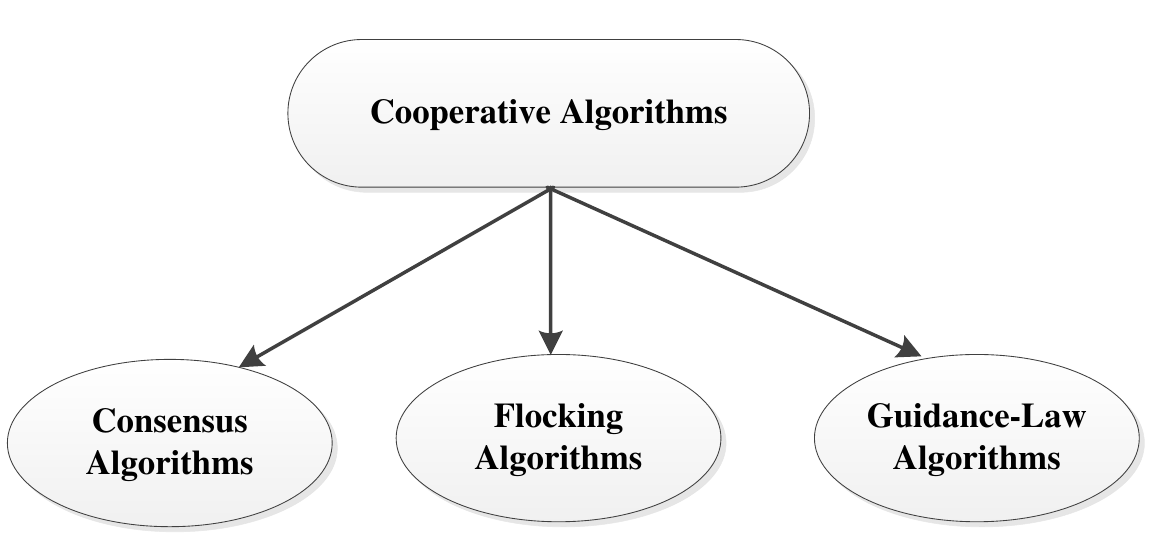} 
\caption {Cooperative algorithms are categorized and explained in three main algorithms.}
\label{F:8}
\end{center}
\end{figure}

The cooperative algorithms can be categorized into three main groups based on their methodologies.  They are 1) Consensus techniques; 2) Flocking techniques; and 3) Formation based techniques. Figure \ref{F:8} shows the main algorithms that are used for the UAVs system. Algorithms for consensus control, flocking control, and formation control are discussed below, respectively. 
\subsection{Consensus Strategies}
In the area of cooperative control, consensus control is an important and complicated problem. In consensus control, a group of agents communicates with each other through a sensing or communication network to reach a common decision. The roots of the consensus control belong to computer science and parallel computing \cite{borkar1982asymptotic,tsitsiklis1986distributed}. In the last decade, the research works of Jadbabaie et al. \cite{jadbabaie2003coordination}  and Olfati-Saber et al. \cite{olfati2004consensus} had a considerable impact on other researchers to work on consensus control problems. Generally, Jadbabaie et al. \cite{jadbabaie2003coordination} provided a theoretical explanation for the alignment behavior of the dynamic model introduced by Vicsek \cite{vicsek1995novel}, and Olfati-Saber introduced a general framework to solve consensus control problem of the networks of the integrators \cite{olfati2004consensus}. In the following subsection, the basic concepts of the consensus control will be explained; then, recent research works in this area will be reviewed. In the cooperative control, the communications among agents are modeled by undirected graphs. Thus, a basic knowledge of graph theory is needed to understand the concept of cooperative algorithms. Therefore, the basic concept of graph theory will be briefly explained, followed by the concept of consensus control theory.
\subsubsection{Graph Theory Basics in Communication Systems}

\begin{figure}
\begin{center}
\includegraphics[scale=0.75]{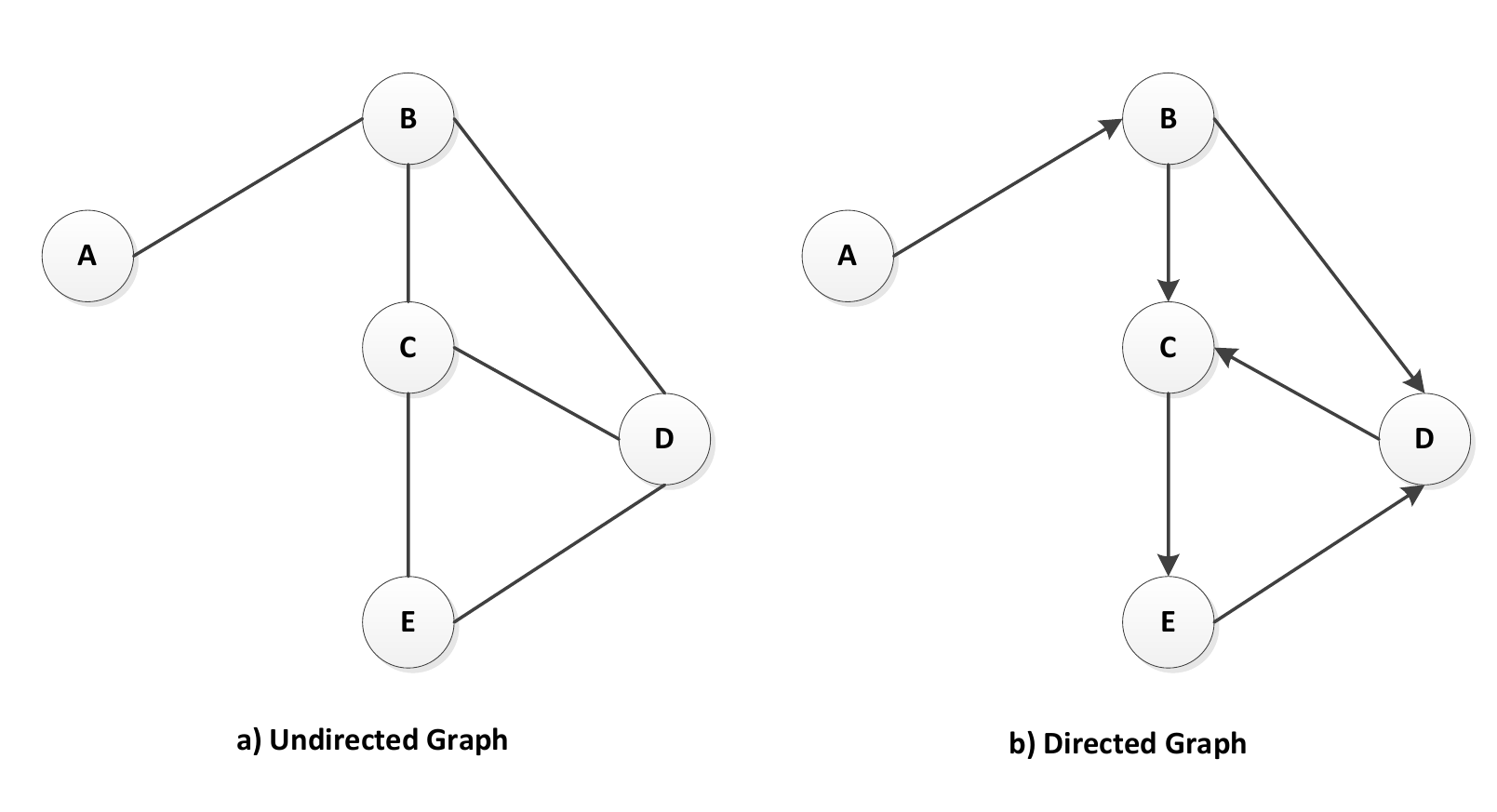} 
\caption {Directed and undirected graph structure.}
\label{F:Graph}
\end{center}
\end{figure}

Communications or sensing among the agents of a team are commonly modeled by undirected graphs. An undirected graph is denoted by
$G=(V,\varepsilon, A)$, where $V=\{1,2,...,N\}$ is the set of $N$ nodes or agents in the network, and $\varepsilon(i,j) \in V\times V$ is set of edges between the ordered pairs of $j^\text{th}$ and $i^\text{th}$ agents. $\beta=[a_{ij}] \in R^{N\times N}$ is the adjacency matrix associated with graph $G$ which is symmetric, and $a_{i,j}$ is a positive value if $(i,j)\in \varepsilon$ and $i\not= j$, otherwise $a_{ij}$ =0 . Figure \ref{F:Graph} shows the basic structure of a directed and undirected graph. \\
For example the adjacency matrix associated with the undirected graph shown in Figure \ref{F:Graph} is:
\begin{equation}
\beta=\begin{bmatrix} 
0 & 1 & 0& 0& 0\\
1 & 0 & 1& 1& 0\\
0 & 1 & 0& 1& 1\\
0 & 0 & 1& 1& 0
\end{bmatrix}
\end{equation}
where node A, B, C, D and E are considered to be nodes 1, 2, 3, 4, and 5 respectively.
\subsubsection{Consensus Control Theory}
The basic concept of the consensus control theory is to stimulate similar dynamics on the state's information of each agent in the group. Based on the communication type, each agent (vehicle) in the system can be modeled based on differential or difference equations.  If the bandwidth of the communication network among the agents is large enough to allow continuous communication, then a differential equation can be used to model agent dynamics. Otherwise, the transmitted data among agents should be sent through discrete packets that need difference equations to model the agent dynamics. These are briefly explained here.\\

\textbf{$\bullet$~ Continuous-time Consensus:} The most common consensus algorithm used for the dynamics defined by differential equations  can be presented as \cite{ren2007information,xie2007consensus,xiao2008asynchronous}
\begin{equation}
\dot{x}_i(t)=-\sum_{j=1}^n a_{ij}(t)\big(x_i(t)-x_j(t)\big),~~~~ i=1,...,n
\label{E:1}
\end{equation}
where $x_i(t)$ is the information state of the $i$th agent, and $a_{ij}(t)$ is the $(i,j)$ element of the adjacency matrix $\beta$ which is obtained from the graph $G$. If $a_{ij}=0$, it indicates that there is no connection between agents $i$ and $j$, subsequently, they cannot exchange any information between them. The consensus algorithm shown in Equation \ref{E:1}, can be rewritten in a matrix form as
\begin{equation}
\dot{x}(t)=-L(t)x(t)
\label{E:2}
\end{equation}
where the Laplacian matrix $L=[l_{ij}]\in R^{N\times N}$ is related to the graph $G$ and can be obtained as follows
\begin{equation}
l_{ij}=\Bigg\{\begin{array}{cc} \sum_{j\in N_i},~~~ i=j\\
-a_{i,j} ,~~~ i\not=j
\end{array}
\label{E:3}
\end{equation}
Since the $l_{ij}$ has zero row sums, an eigenvalue of $L$ is $0$, which is associated with an eigenvector of $1$. Because $L$ is symmetric, in a connected graph, $L$ has $N-1$ real eigenvalues on the right side of the imaginary plane. Thus, $N$ eigenvalues of $L$ can be defined as follows\\
\begin{equation}
0=\lambda_1<\lambda_2\leq \lambda_2...\leq \lambda_N
\label{E:4}
\end{equation}

Based on this condition and the fact that $L$ is symmetric,  the diagonalized $L$ can be obtained by orthogonal transformation matrix as
\begin{equation}
L=PJP^T
\label{E:5}
\end{equation}
where $P$ consists of the eigenvectors of the $L$ and $J$ is a diagonal form of $L$ which are defined as follows\\
\begin{center}
$P=[r_1~ r_2~.~.~.~ r_n]$\\
\end{center}
\begin{center}
$J=\begin{bmatrix}
0& 0_{1\times (N-1)}\\
0_{(N-1)\times 1} & \gamma
\end{bmatrix}$
\end{center}
where $\gamma$ is a matrix with diagonal form which contains $N-1$ eigenvalues of $L$ which have positive values, and $r_i$, $i\in \{1,2,...,N\}$ describes the eigenvectors of $L$ where $r_i^T r_i=1$ \cite{ren2007information}.\\
It can be claimed that \textit{consensus} is achieved for a team of agents for all $x_i(0)$ and all $i,j=1,.~.~.~, n$, if  $lim _{t\rightarrow \infty}|x_i(t)-x_j(t)|=0$ \cite{ren2007information}.

\textbf{$\bullet$ ~Discrete-time Consensus:}
The discrete-time consensus is used when the communication bandwidth among the agents in the team is weak or occurs at discrete instants. In this case, the information states are updated through difference equations. The following form commonly presents the discrete-time consensus \cite{chen2012event, zhang2017data,ding2015event,jin2017collision}
\begin{equation}
x_i[k+1]=\sum_{j=1}^n d_{ij}[k] x_j[k], ~~~ i=1,~.~.~.~,n
\label{E:6}
\end{equation}
where $k$ is the solving step associated to the communication event; $d_{ij}[k]$ is the $(i,j)$ element of the stochastic matrix $D=[d_{ij}]\in R^{n\times n}$.\\
The discrete time consensus algorithm in Equation \ref{E:6} can be rewritten in a matrix form as
\begin{equation}
x[k+1]=D[k]x[k]
\label{E:7}
\end{equation}
where $D=[d_{ij}]>0$, if $i\not=j$ and the information flows from the agent $j$ to $i$, otherwise $d_{ij}[k]=0$ \cite{ding2015event}.\\
Similarly, a discrete-time \textit{consensus} is achieved  for a team of agents for all $x_i[0]$ and all $i,j=1,.~.~.~, n$, if  we have $lim _{k\rightarrow \infty}|x_i[k]-x_j[k]|=0$ \cite{ding2015event}.

\subsubsection{Consensus Recent Researches}
\begin{figure} [!h]
\begin{center}
\includegraphics[scale=0.4]{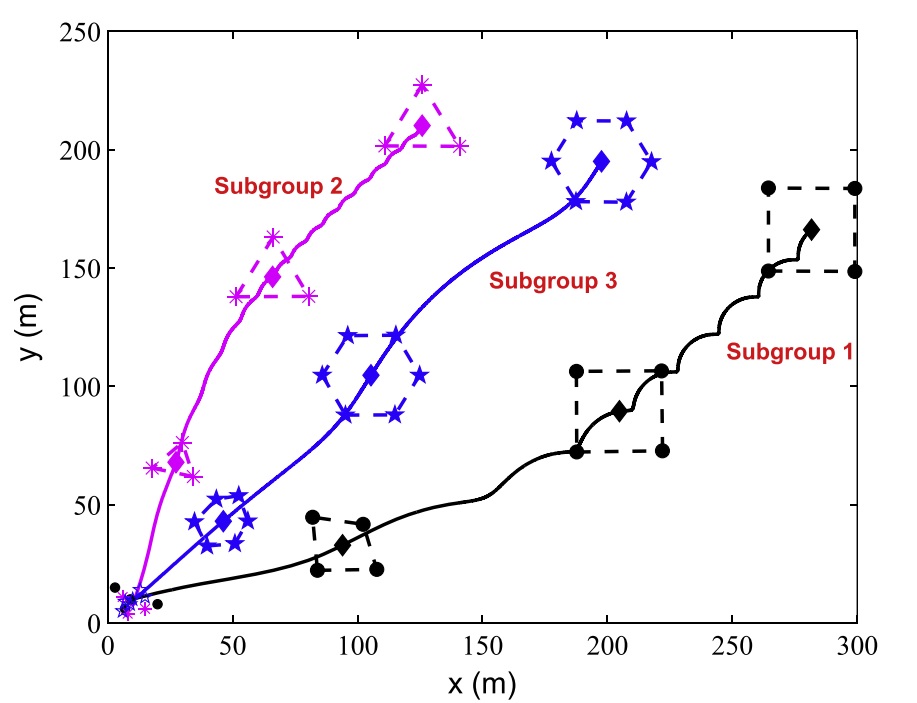} 
\caption {Multi-consensus control of three subgroups by Han et al. \cite{han2017multi}}
\label{F:10}
\end{center}
\end{figure}

The consensus control algorithm, which is based on graph theory, has received a growing interest among researchers \cite{jamshidi2011cyber,rezaee2015average}.
Jamshidi et al. developed a testbed and a consensus technique for cooperative control of UAVs \cite{jamshidi2011cyber}. Rezaee and Abdollahi proposed a consensus protocol for a class of high-order multi-agent systems \cite{rezaee2015average}. They showed how agents achieve consensus on the average of any shared quantities using their relative positions. Li presented a geometric decomposition approach for cooperative agents \cite{li2016unified}. Under topology adjustments, decomposing a system into sufficiently simple sub-systems facilitates subsequent analyses and provides the flexibility of choice.
Liang et al. introduced an observer-based discrete consensus control system. The nonlinear observer was used to obtain the states of the agents, and a feedback control law was designed based on the data received from the observer \cite{liang2015distributed}.
Xia et al. introduced an optimal design for consensus control of agents with double-integrator dynamics with collision avoidance considerations \cite{xia2016formation}. Han et al. introduced a nonlinear multi-consensus control strategy for multi-agent systems \cite{han2017multi}. In their research, both of the switching and fixed topology were considered, and their consensus controller could control three subgroups, as shown in Figure \ref{F:10}. They were also compared their research work with their previous work \cite{han2013multiconsensus} in which they could reduce the convergence time in consensus control. Shoja et al. introduced an estimator based consensus control scheme for agents with nonlinear and nonidentical dynamic systems \cite{shoja2017surrounding}. In their design, they used an undirected graph model for their communication system among the agents, and multiple leaders were considered in their design. 
A sliding mode consensus control design for double-integrator multi-agent systems and 3-DoF helicopters was introduced by Hou et al. \cite{hou2016finite}. The advantage of their proposed method was achieving synchronization in the presence of disturbances and the ability to be implemented on 3-DoF model of helicopters.\\
Taheri et al. introduced an adaptive fuzzy wavelet network approach for consensus control of a class of a nonlinear second-order multi-agent system \cite{taheri2017adaptive}. The adaptive laws were obtained using the Lyapunov theory to maintain the nonlinear dynamic stability. Then, an adaptive fuzzy wavelet network was used to compensate for the effect of unknown dynamics and time delay in the system. However, the authors didn't address the design of a consensus control design for a second-order multi-agent system with a directed graph.
Neural networks and robust control techniques have been used in \cite{zhang2012adaptive} and \cite{el2014neuro} to design a  consensus controller for higher-order multi-agent systems and their semi-global boundedness of consensus error was ensured by choosing sufficiently large control gains.  
Consensus Fault Tolerant Controllers (FTC) with the ability to tolerate faults in the actuators of agents in a multi-agent system were also investigated \cite{gallehdari2016distributed,gallehdari2017distributed,hua2017distributed}. Gallehdari et al. introduced an online redistributed control reconfiguration approach that employed the nearest neighbor information and the internal Fault Detection and Identification (FDI) of the agent to keep the consensus control in the presence of faults in the actuators. They used the first-order dynamic model for their agents, and their proposed controller was designed based on minimizing the cost of faulty agent performance index which led to optimizing the performance index of the team. Later, they developed their work to optimize all the agents in the consensus FTC system \cite{gallehdari2017distributed}. 
Hua et al. introduced a consensus  FTC design for time-varying high-order linear systems which could tolerate faults in the actuators \cite{hua2017distributed}.\\
Wang et al.introduced a new smooth function-based adaptive consensus control approach for multi-agent systems with nonlinear dynamic, unknown parameters and uncertain disturbances without the need for the assumption of linearly parametrized reference trajectory \cite{wang2016distributed}. Their approach was based on the premise of transmitting data among the agents based on an undirected graph model. Later, they extended their work for directed graph model as well \cite{wang2017distributed}.

\subsection{Flocking Based Strategies}

Flocking can be defined as a form of collective behavior of a group of interacting agents with mutual objectives. Flocking algorithms are inspired by a flock of birds and developed based on Reynolds rules. Reynolds modeled the steering behavior of each agent based on the positions and velocities of nearby flock-mates, using three terms of separation (collision avoidance), alignment (velocity matching), and cohesion (flock centering) \cite{reynolds1987flocks}.\\

\subsubsection{Flocking Control Theory}
Similar to consensus algorithms, flocking algorithms are based on graph theory. Unlike the formation strategies that require the group of agents to be in a particular shape, the group of agents in the flocking is not necessarily in a rigid shape or form. In other words,  in flocking control, as long as the flock goals are satisfied, transition in the shape of the flock is allowed, e.g., it can be transformed from a rectangular shape to a triangular shape.\\
Several flocking algorithms have been devised for multi-agent systems with a second-order dynamic model \cite{moshtagh2007distributed,olfati2006flocking,saif2014flocking}. The following equation of motion can present a group of agents with a second-order dynamic model.\\
\begin{equation}
\Bigg\{ \begin{array}{cc}
\dot{q}_i=p_i\\
\dot{p}_i=u_i
\end{array}
\label{E:8}
\end{equation}
where $q_i$ is the position of agent (node) $i$ and  $p_i$ is the velocity. $p_i, q_i, u_i \in R^m$ and $i\in V={1,2,...,N}$ (set of N nodes or agents in the network). Flocking algorithms consists of three terms: 1) a gradient-based term, 2) a consensus term, and 3) a navigational feedback term, and can be presented as follows \cite{olfati2006flocking}
\begin{equation}
u_i=\underbrace{\sum_{j\in N_i}\phi_{\alpha}(\|q_j-q_i\|)n_{ij}}_{gradient-based~ term}+\underbrace{\sum_{j\in N_i}a_{ij}(q)(p_j-p_i)}_{consensus~term}+\underbrace{f_i^{\gamma}(q_i,p_i,q_r,p_r)}_{Navigational-based~term}
\label{E:9}
\end{equation}
where $\phi(\bullet)$ is a potential function, and $n_{ij}=\sigma_{\epsilon}(q_j-q_i)=(q_j-q_i/\sqrt[]{1+\epsilon\|q_jq_i\|^2})$ is a vector along the line connecting $q_i$ to $q_j$ in which $\epsilon\in(0,1)$ is a constant parameter of the norm in $\sigma$-norm. The pair $(p_r,q_r)\in R^m\times R^m$ is the state of a $\gamma$ agent. The navigational feedback term $f_i^{\gamma}$ is given as follows
\begin{equation}
f_i^{\gamma}(q_i,p_i,q_r,p_r)= -c_1(q_i-q_r)-c_2(p_i-p_r),~~~c_1,c_2>0
\label{E:10}
\end{equation}
The flocking algorithm in Equation \ref{E:9} can be developed by using some updating terms to tackle the problem of uncertainties in the flock control.
One major problem with flocking control is its incapability of covering a large area. Thus, a semi-flocking algorithm was introduced to tackle this problem \cite{semnani2015semi}.  In the semi-flocking algorithm, the navigation feedback term is modified to make each agent able to decide whether to track a target or to search for a new one.\\
\subsubsection{Flocking Recent Researches}

Moshtagh and Jadbabaie introduced a novel flocking and velocity alignment algorithm to control the kinematic agents using graph theory \cite{moshtagh2007distributed}. In their design which was capable of flocking control in two and three dimensions, they used a geodesic control to minimize the misalignment potential which leads to flocking and velocity alignment. They also demonstrated that their method could keep the flocking even when the topology of proximity graph changes, and as long as the joint connectivity is well-maintained, the algorithm will be successful in consensus control. However, to guarantee the flocking success, still, one problem has to be solved, and that is how to keep the connectivity condition in the proximity graph.
Olfati-Saber introduced a systematic approach for the generation of cost functions for flocking \cite{olfati2006flocking}. In these cost functions, the deviation from flock objects will be penalized. They demonstrated that a peer-to-peer network of agents could be used for the migration of flocks and the need for a single leader for the flock can be eliminated. The simulation results for flocking hundreds of agents in 2-D and 3-D, squeezing, and split/reuniting maneuvers were provided that showed the success of the proposed algorithm in the presence of obstacles. Saif et al. introduced a Linear Quadratic Regulator (LQR) controller for a flock of UAVs which is independent of the number of agents in the flock \cite{saif2014flocking}. This control strategy can satisfy the Reynolds rules, and independent of the number of UAVs in the flock it allows designing an LQR controller for each of the UAVs. Chapman and Mesbahi designed an optimal controller for UAV flocking in the presence of wind gusts, using a consensus-based leader-follower system to improve velocity tracking \cite{chapman2011uav}.\\
Tanner et al. introduced a control law for flocking of multi-agent systems with double integrator dynamics and arbitrary switching in the topology of agent interaction network \cite{tanner2007flocking}. The non-smooth analysis was used to accommodate arbitrary switching the agent's network, and they demonstrated that their control law is robust against arbitrary changes in the agent communication network as long as they are connected in their maneuvers.
Hung and Givigi developed a model-free reinforcement learning approach to flocking of small fixed-wing UAVs in a leader-follower topology \cite{hung2017q}. In their study, agents experience disturbances in a stochastic environment. The advantage of their online learning design is that their model is not dependent on the environment; hence, it can be implemented in a different environment without any information about the plant and disturbances in the system. This characteristic increases the adaptability of the system to unforeseen situations. However, the learning rate and convergence speed of flocking are two factors that still need to be solved.
Quintero et al. introduced a leader-follower design for flocking control of multiple UAVs to conduct a sensing task \cite{quintero2013flocking}. The UAVs were considered as fixed-wing airplanes flying at a constant speed with fixed altitude which limits its movement in a 2-D planar surface. In their strategy, each of the followers is controlled using a stochastic optimal control problem where the cost function is the heading and distance toward the leader. This algorithm was successfully applied and implemented in three UAVs equipped with cameras; however, the offline solving the optimization problem cannot guarantee the flocking behavior of the system in the presence of nonlinear behavior of flock and its agents.\\ 
McCune et al. introduced a framework based on a Dynamic Data-Driven Application System (DDDAS) to predict, control, and improve decision making artificial swarms using repeated simulations and synergistic feedback loops \cite{mccune2014control}.  Using this strategy helps to improve the decision making in the process of swarm control; however, the time frame for the real-time application of this strategy has not been considered which can affect the effectiveness of this approach.
Martin et al. \cite{martin2014multiagent} considered a system of agents with second-order dynamics. They determined conditions to ensure that agents agree on a common velocity to achieve system flocking. The significance of their design was the allowance for disconnected communication links that were unnecessary for flocking. Practical bounds for two different communication rules were investigated; first, the agents communicate within the radius of communication bound, second, agents communicate with each other with different and randomly communication radiuses. Overall, they concluded that by choosing a proper initial velocity disagreement or by setting a small enough time step, flocking can be achieved with random communication radiuses. One of the drawbacks of their approach was an asymmetric requirement in the interaction among the agents. Generally, other types of interaction (i.e., assuming that agents interact with the nearest neighbors with a fixed parameter which is called topological interaction rule) can happen in the flock.
Riehl et al. introduced a receding-horizon search algorithm for cooperative UAVs \cite{riehl2011cooperative}. In order to find a target in the minimum time, each of the UAVs was equipped with a gimbal sensor which could be rotated to observe the nearby target; then by gathering information on a potential location for the target, they could find it. The algorithm helps to minimize the expected time for finding the target by controlling the position of UAVs and their sensors. The optimization process is a receding horizon algorithm based on a graph with variable target Probability Density Function (PDF). This algorithm was successfully tested using two small UAVs equipped with gimbaled video cameras.\\
\subsection{Guidance-law based cooperative control}
\begin{figure} [!h]
\begin{center}
\includegraphics[scale=0.4]{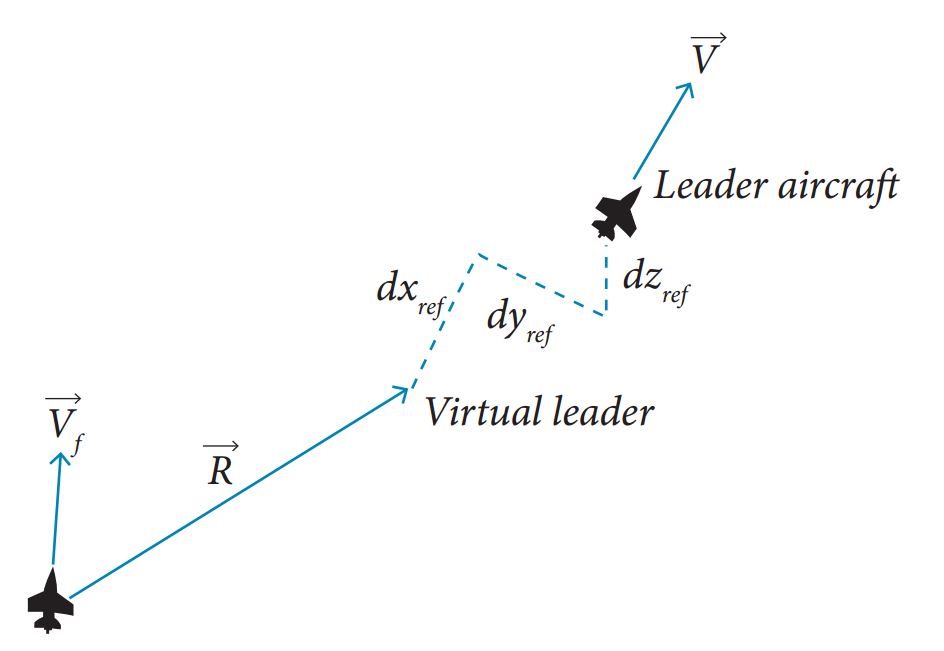} 
\caption {Geometry of the PP guidance algorithm \cite{shneydor1998missile}}
\label{F:11}
\end{center}
\end{figure}

This subsection is separated from the other cooperative control techniques because they do not deal with the guidance system in their design. In order to achieve a formation, the acceleration and angular velocity of each agent in the formation group should be calculated separately \cite{naeem2003review}. To this aim, guidance law techniques are used to obtain the desired acceleration and angular velocities. Pure pursuit (PP) guidance algorithm is one of the most practical leader-follower guidance techniques in the formation control. This algorithm was initially implemented on ground-attack missile systems that aim to hit the target \cite{lin1991modern}. Later by introducing the concept of the virtual leader (or target) it has been developed for the formation of flight control which the followers keep their Line of Sight (LoS) in-line with the leader movement. In other words, the velocity direction of the agents should be aligned with the velocity of the leader \cite{naeem2003review}.\\
In the PP algorithm, between the follower speed vector $\vec{V}$ and the virtual leader $\vec{R}$ the following equation is maintained:
\begin{equation}
\vec{V_f}\times \vec{R}=0
\end{equation}
Figure \ref{F:11} shows the geometry between the virtual leader and the follower in the PP algorithm. In this figure, $d_{x_{ref}}$, $d_{y_{ref}}$, and  $d_{z_{ref}}$ represent the distance between the leader and the virtual leader in the longitudinal axis,  lateral axis, and the vertical axis, respectively. The required acceleration in the follower aircraft to reach the virtual leader can be calculated as follows \cite{shneydor1998missile,yamasaki2009advanced}
\begin{equation}
\vec{A}_f=\frac{N(\vec{V_f}\times \vec{R})\times\vec{V_f}}{\|\vec{V_f}\|~ \|\vec{R}\|}
\end{equation}
where $N$ is the navigational constant which is usually chosen between 0.3 to 0.5.
Proportional navigation (PN) guidance is another candidate that can be applied in the formation control design; however, because when the closing velocity is negative (the leader velocity is higher than the follower aircraft), the PN guidance is likely to guide the follower away from the leader \cite{yamasaki2009advanced}. In contrast, the PP guidance does not depend on the leader velocity and always guides the follower in the direction of the leader. Thus, we discussed the PP guidance laws application in the control design of the flight formation systems.

\subsubsection{Guidance-law based recent researches}
 
Gu et al. \cite{gu2006design} introduced a nonlinear leader-follower based formation control law. A two-loop controller was designed where nonlinear dynamic inversion (DI) was used to design the velocity and position tracker in the outer-loop, and a linear controller was used to track the leader attitude in the inner-loop. This two-loop design is based on the difference in the changing rate of the inner-loop and outer-loop dynamic parameters. The introduced controller was experimentally tested on two WVU YF-22 aircrafts as leader and follower. The experimental results demonstrated the effectiveness of their proposed formation control law. 
Yamasaki et al. introduced a PP-guidance based formation control system for a group of UAVs \cite{yamasaki2009advanced}. Their proposed control system uses a PP guidance algorithm and a velocity controller based on the DI control technique to avoid a collision. The attitude controller of the follower aircraft was designed based on a two-loop DI controller. Sadeghi et al. improve the Yamasaki work \cite{yamasaki2009advanced} and introduced a new approach to integrating the guidance and control system through a PID control design \cite{sadeghi2015novel}. Their proposed approach could improve the PP guidance algorithm accuracy and the maneuverability of the formation group.   

Zhu et al. introduced a least-squares method for the estimation of the leader location, then, a guidance law based on sliding mode control was designed to control the heading rate of the follower aircrafts toward the leader estimated location \cite{zhu2014cooperative}. Ali et al. presented a  guidance law for lateral formation control of UAVs based on sliding mode theory \cite{ali2016lateral}. Two sliding surfaces were integrated into series to improve the control response in the formation design. 
A new approach for UAVs formation control considering obstacle/collision avoidance using modified Grossberg neural network (GNN) was developed by Wang et al.\cite{wang2007cooperative}. In order to track the desired trajectory, a model predictive controller was used. They simulated their collision/obstacle avoidance design in a 3-D environment.
A LOS guidance law approach for formation control of a group of under-actuated vessels is studied in \cite{borhaug2011straight}. In their approach, a nonlinear synchronization controller was combined with the LOS-based path following controller to make the overall system more robust and controllable under the under-actuation situation.

\section{Summary and Conclusion}
In this chapter, the algorithms and applications of cooperative control techniques for UAVs are reviewed. By categorizing the recent researches to applications and methods, each was discussed separately. The latest studies in the field of cooperative control of UAVs have been investigated and the advantages and disadvantages of methods were discussed. Applications of cooperative UAVs mission in various fields have been explored. Although some studies in the cooperative field may have been missed in this survey, it is hoped that this survey would be helpful for researchers to overview the major achievements in cooperative control of UAVs.

\section*{Acknowledgement}
The first and second author would like to thank Dr. Kang Yen for his guidance and effort toward this paper.

\bibliography{book_chapter.bib}
\bibliographystyle{ieeetr}

\end{document}